\documentclass{article}

\newcommand{\RRonly}[1]{}
\newcommand{\SoCG}[1]{#1}     
\newcommand{\SoCGsubmit}[1]{}

\SoCG{
}

\RRonly{\usepackage{RR}}
\SoCGsubmit{
\voffset=-2cm
\hoffset=-1cm
\textwidth              15cm
\textheight             24cm

}

\newtheorem{theorem}{Theorem}
\newtheorem{lemma}[theorem]{Lemma}
\newenvironment{proof}{
   \begin{list}{}{\addtolength{\leftmargin}{-\parindent}
   \addtolength{\rightmargin}{0.125in}}
   \item[]
   {\bf Proof\ }}
 {\hspace*{0.01mm}\hfill{\vspace{1ex}\rule[.012in]{.07in}{.07in}}
   \end{list}
}

\begin{document}

\SoCG{
\title{Improved Incremental Randomized Delaunay Triangulation.\thanks{
      \scriptsize
        This work was partially supported by ESPRIT LTR 21957 (CGAL)}}
\author{Olivier Devillers\thanks{
      \scriptsize
       INRIA, BP93, 06902 Sophia Antipolis.
        Olivier.Devillers@sophia.inria.fr.
}}
\date{}
\maketitle
}

\section{Introduction}

The computation of the Delaunay triangulation of
a set of $n$ points in the plane is one of the classical
problems in computational geometry
and plenty of algorithms have been proposed to solve it.

These Delaunay algorithms can have different characteristics:
\begin{itemize}
\item Optimal on worst case data, i.e. $O(n\log n)$ time.
\item Good complexity on random data only
\item Randomized
\item On-line vs off-line
\end{itemize} 

In the current trade-off between algorithmic simplicity,
practical efficiency and theoretical optimality,
practitioners often choose the simplicity and practical efficiency
taking the risk of having  bad performance
on some special kind of data.






Our aim  is to conciliate many of the above aspects,
namely to obtain an incremental algorithm using simple
data structure having good practical performance
on realistic input and still provable
 $O(n\log n)$ computation time on any  data set.

{\bf Previous related work}

Our work is strongly related to some previous algorithms
for Delaunay triangulation. All these algorithms are incremental
and their complexity is randomized, they use some
location structure to find where the new point is inserted,
and then update the triangulation.

The first idea of a randomized incremental construction
for the Delaunay triangulation \cite{bt-hrodt-86}
uses a location structure based on the history of the Delaunay
triangulation: the Delaunay tree.
Point $p_i$ is inserted at time $i$, and to find where point
$p_n$ fell, $p_n$ is located in all the triangulations at times
$1$ to $n-1$; the location at time $i+1$ is
deduced from the location at time $i$.
This  idea  yields an expected $O(n\log n)$ complexity
\cite{bt-rcdt-93,gks-ricdv-92} if the points are inserted in a
random order.
The drawbacks of this approach are the following:
the location structure consists of the history of the construction
and thus strongly depends on the insertion order, and the
additional memory needed cannot be controlled.
(The expected memory is proved to be $O(n)$ and is experimentally about
twice the size of the final triangulation.)

Mulmuley \cite{m-rmstd-91} proposed a location structure independent
of the insertion order. The structure has $O(\log n)$ levels,
each level being a random sample of the level below.
At each level, the Delaunay triangulation of the points is computed,
and the overlapping triangles at different levels are linked to enable
location of new points.
This structure has the advantage of being independent of
the order of insertion, of ensuring an $O(\log^2n)$ location
time for any point, and of allowing deletions in an easier
way than the Delaunay tree \cite{dmt-fddtl-92}.
However, the additional memory is still important and the
location structure is not especially simple.

In 1996, M\"ucke, Saias and Zhu \cite{msz-frplw-96}
proposed a very simple structure to handle triangulation
of random points.
The structure  reduces to a random subset of
$\sqrt[3]{n}$ points, and pointers from these points to an
incident triangle in the Delaunay triangulation.
A new point is located by finding the nearest neighbor
in the sample by  brute force, and walking in the triangulation.
For  evenly distributed points, the expected complexity
of the algorithm is $O(n^{\frac{4}{3}})$ with a small constant,
which makes it competitive with many $O(n\log n)$ algorithms.
But  for some
data (for example points on a para\-bola) the complexity
increases to $O(n^{\frac{5}{3}})$.

{\bf Overview}

Our approach uses a structure with levels similar to Mulmuley, but
with   simple relations between levels.
This allows better control of the memory overhead.
The transition between two levels is not direct
as in Mulmuley, but
 uses a march 
 similar M\"ucke, Saias and Zhu
to locate point in triangulations.

In Section \ref{algo} we present the algorithm,
in Section \ref{analysis} we prove that the expected complexity
of constructing the Delaunay triangulation is $O(n\log n)$.
The parameters of the data structure are then tuned to
minimize the constant in the case of random points and are shown to yield
an excellent behavior in Section \ref{tune},
we pay special attention to the comparison with
the method of M\"ucke, Saias and Zhu.
Finally we give some implementation remarks and practical results
in Section \ref{implement}.

\section{Algorithm\label{algo}}

Let  ${\cal S}$ be a set of $n$ sites in the plane.
The aim is to compute the Delaunay triangulation
${\cal DT}_{\cal S}$ of ${\cal S}$ and to maintain it efficiently
under insertions and deletions.

\subsection{The location structure}

The algorithm uses a data structure composed of different levels.
Level $i$ contains the Delaunay triangulation ${\cal DT}_i$ of 
a set of sites ${\cal S}_i$.

The sets ${\cal S}_i$ forms a decreasing sequence of random
subsets of  ${\cal S}$ based on a Bernoulli sampling technique \cite{mr-ra-95,m-cgitr-93}:
\[
{\cal S} = {\cal S}_0 \supseteq {\cal S}_1  \supseteq {\cal S}_2 \supseteq\ldots\supseteq {\cal S}_{k-1} \supseteq {\cal S}_k
\] \[
Prob(p\in {\cal S}_{i+1} \; |\; p\in{\cal S}_i ) = \frac{1}{\alpha}\in]0,1[.
\]

The data structure is fairly simple: it contains the points of ${\cal S}$
and the triangles of all the triangulations ${\cal DT}_i$.
A point $p\in{\cal S}$ such that $p\in{\cal S}_i\subseteq\ldots\subseteq{\cal S}_0$ and $p\not\in{\cal S}_{i+1}$
is said to be a {\em vertex of level $i$} and
 has a link to a Delaunay triangle of ${\cal DT}_j$
incident to  $p$ for all $j$ for $0\leq j\leq i$.
A triangle of ${\cal DT}_i$ has links to its three neighbors in ${\cal DT}_i$
and to its three vertices.The number $k$ of levels is not fixed; for each
point random trials  decide its level, and the point with highest level
determines $k$.

\subsection{Location of a query}

For the location of a query $q$, we start
at a known vertex $v_{k+1}$ of the highest level $k$.
Then we search for $v_k$, the vertex of ${\cal DT}_k$ nearest to $q$.
Since $v_k$ is also a vertex of ${\cal DT}_{k-1}$, we search for
$v_{k-1}$, the nearest neighbor of $q$ in ${\cal DT}_{k-1}$,
starting at $v_k$.
The search is continued descending the different levels.
At each level $i$,
the nearest vertex $v_i$  of $q$ in ${\cal DT}_i$ is determined.

\begin{figure} \begin{center} 
\def\IPEfile{FindNN.ipe}\input{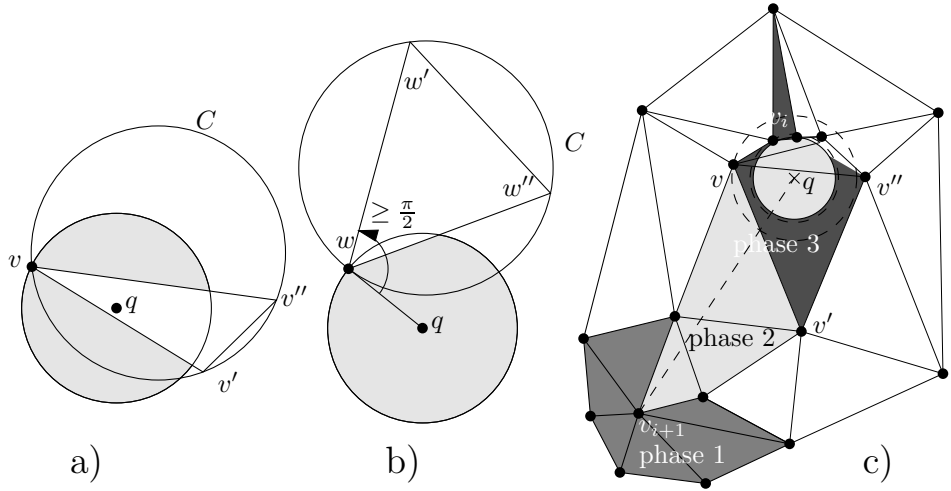} 
\caption{\label{FindNN}Search for $v_i$.}
\end{center} \end{figure}

At level $i$ the search of $v_i$ is carried out in three phases:
\begin{itemize}
\item  First phase: from $v_{i+1}$, we have a link to a triangle of  ${\cal DT}_i$ 
having $v_{i+1}$ as vertex. All triangles incident to
$v_{i+1}$ are explored to find the triangle containing the segment $v_{i+1}q$.
\item Second phase: all the triangles of ${\cal DT}_i$ intersected by $v_{i+1}q$
are visited, walking along the segment $v_{i+1}q$ up to the triangle  $t_i$
that contains $q$.
\item Third phase: using neighborhood relationships between triangles,
we will traverse few triangles of ${\cal DT}_i$ from $t_i$ to find $v_i$.
If $vv'v''$ are the three vertices of $t_i$, and, without loss of generality, $v$ is closer
to $q$ than $v'$ and $v''$, then $v_i$ is either $v$ or it lies
in the disk of center $q$ and passing through $v$ (shaded on Figure \ref{FindNN}a);
thus the search for  $v_i$ has to be done only in the direction of the neighbors
of $t_i$ through the edges $vv'$ and $vv''$ and the
neighbor through the edge $v'v''$ can be ignored
(the portion of the shaded disk in that direction is inside the disk
through $vv'v''$ which is empty) .
For each such triangle, the distance to the new vertex is computed and the
algorithm  maintains  the closest visited vertex.
For a visited triangle $ww'w''$ such that $w$ is the nearest to $q$ among $ww'w''$
the neighbor triangle through edge $ww'$ (resp $ww''$) will be visited
if angle $qww'$ is smaller than $\frac{\pi}{2}$ (Figure \ref{FindNN}b).
\end{itemize} 
Figure \ref{FindNN}c show the triangles visited by the different phases of the search. 

\subsection{Updates}

Because of its simplicity, the data structure is fairly easy to update.
Maintaining it dynamically provides a fully dynamic triangulation algorithm.
The links between the different levels do not use any complicated data
structure simply vertices know a triangle
at all levels in which they appear.

To delete a point from ${\cal S}$, just delete the
corresponding vertex at all the levels where it appears,
which can be done in  time sensitive to $d$ the degree of
that vertex.
On average $d=6$ and thus some of the following algorithms can be used.
A complicated algorithm \cite{agss-ltacv-89} of deterministic complexity $O(d)$,
a simple randomized $O(d)$ algorithm \cite{c-bvdcp-86}
can be used
or simpler solutions of complexity $O(d\log d)$ or even $O(n^2)$
may be good in practice.

Inserting a point in  ${\cal S}$ reduces to
locating the new point at all levels,
computing its level $i$ 
and inserting  the new vertex at all levels
$j,0\leq j\leq i$ (which is  sensitive to the
degree of the new vertex once the location is done).
The insertion using the standard algorithm \cite{l-scsi-77}.

\section{Worst-case~randomized \mbox{analysis}\label{analysis}}
The  analysis will rely on the randomization
in the construction of the random subsets ${\cal S}_i$
and the points of ${\cal S}$ are assumed to be inserted in a random order.
In this section, no assumption applies to the data distribution,
which can be in the worst case.
As usual in theoretical computational geometry, we make only an
asymptotic analysis and give rough upper bounds for the constants.
In the next section, parameter $\alpha$ will be tuned to get
a tight constant in the special case of evenly-distributed points.

Let ${\cal S}$ be a set of $n$ points organized in the structure described in Section \ref{algo}
and $q$ a point to be inserted in ${\cal S}$.
Since we have assumed a random insertion order, $q$ is a random point of
 ${\cal S}\cup\{q\}$.

We denote $n_i = |{\cal S}_{i}|$ and
${\cal R}_i={\cal S}_{i}\cup\{q\}$.

Notice that, thanks to the random insertion order,
${\cal R}_i$ is a random subset of size $n_i+1$
of ${\cal R}_{i-1}$ and
$q$ is a random element of ${\cal R}_i$.

The cost of exploring all the triangles incident to $v_{i+1}$
at the first  phase of the march of level $i$
is the degree of $v_{i+1}$ in ${\cal DT}_i$.
The cost of the second phase is the number
of triangles intersected by segment $v_{i+1}q$.
The cost of the third phase is the number of candidate vertices
visited during the search of $v_i$ from $t_i$.

\begin{lemma} \label{ana_phase1}
The expected degree of $v_{i}$ in ${\cal DT}_{i-1}$ is $O(1)$.
\end{lemma}
\begin{proof}
Let ${\cal NN}$ be the nearest neighbor graph of ${\cal R}_i$:
that is,
the vertices of ${\cal NN}$ are the points of ${\cal R}_i$, and
$q,v\in{\cal R}_i$ define an edge of ${\cal NN}$ if
and only if $v$ is the nearest neighbor of $q$ (denoted by $v=NN(q)$)
or  $q$ is the nearest neighbor of $v$  in ${\cal R}_i$.
${\cal NN}$ is well known to be a subgraph of
${\cal DT}_{{\cal R}_i}$, the Delaunay triangulation of ${\cal R}_i$,
and to have maximum degree 6 \cite{py-nng-92}.

We denote by $d^{\circ}_{{\cal DT}_{i-1}}(v)$
the degree of $v$ in ${\cal DT}_{i-1}$,
and by $E_{v\in {\cal R}_i\subset\{q\}}$ the expectation when $v$ is
chosen uniformly in ${\cal R}_i\subset\{q\}$. 
Then we have
\[
E_{v\in {\cal R}_i\subset\{q\}} \left(d^{\circ}_{{\cal DT}_{i-1}}(v)\right)
= E_{v\in {\cal R}_{i-1}\subset\{q\}} \left(d^{\circ}_{{\cal DT}_{i-1}}(v)\right)
< 6
\]
notice that $d^{\circ}_{{\cal DT}_{i-1}}(v)$ is a random variable;
result holds
since ${\cal R}_i$ and ${\cal R}_{i-1}\subset\{q\}$ are random subsets of
 ${\cal R}_{i-1}$
and that the average degree of a vertex in a triangulation is less than 6.

But even if $q$ is a random point in ${\cal R}_i$, the vertex
$v_i$, the nearest neighbor of $q$
in ${\cal R}_i$, is not uniformly random.

\begin{eqnarray*}
E_{q\in {\cal R}_i} \left(d^{\circ}_{{\cal DT}_{i-1}}(NN(q))\right)
&=& E \left(\frac{1}{|{\cal R}_i|}\sum_{q\in {\cal R}_i}  d^{\circ}_{{\cal DT}_{i-1}}(NN(q))\right)\\
&=& \frac{1}{|{\cal R}_i|}E \left( \sum_{v\in {\cal R}_i}\sum_{\;q\in\{\rho;v=NN(\rho)\}} d^{\circ}_{{\cal DT}_{i-1}}(v)\right)\\
&<& \frac{1}{|{\cal R}_i|}  E \left(\sum_{v\in {\cal R}_i} 6 d^{\circ}_{{\cal DT}_{i-1}}(v)\right)\\
&\leq& 36
\end{eqnarray*} 
\end{proof} 

\begin{lemma} \label{number_w2}
Given $w\in{\cal R}_i$,
the expected number of vertices $q$ of ${\cal R}_i$ such that
$w$ belongs to the disk of center
$q$ and passing through the nearest neighbor of $q$ in ${\cal R}_{i+1}$
is less than ${6}{\alpha}$.
\end{lemma}
\begin{proof}
Let $w\in{\cal R}_i$ and let $w=q_0,q_1,q_2\ldots q_k$
be the points of ${\cal R}_i$ lying in a section of angle $\frac{\pi}{3}$
 having apex $w$ sorted by increasing
distance to $w$.
Clearly, a disk of center $q_l$ passing through $q_j$ ($j<l$) cannot contain $w$
and thus, if $q=q_l$, a necessary condition for $w$ to be in the disk
having as  diameter the segment 
defined by $q$ and the nearest neighbor of $q$ in ${\cal R}_{i+1}$
is that no point of $\{q_0,\ldots q_{l-1}\}$ is in
 the sample ${\cal R}_{i+1}$ which has
probability $(1-\frac{1}{\alpha})^l$.

Using six sections around $w$ to cover the whole plane,
and summing over the choice of $q\in{\cal R}_i$ we get the
claimed result.
Notice that the disk of center
$q$ and passing through the nearest neighbor of $q$
contain the disk of diameter  the line segment defined by these
two points, and thus the bound apply also to that circle.
\end{proof}

\begin{lemma} \label{ana_phase2}
The expected number of edges of ${\cal DT}_i$ intersecting segment
$qv_{i+1}$ is $O({\alpha})$.
\end{lemma}
\begin{proof}
Let $e$ be an edge of ${\cal DT}_i$ intersecting
segment $qv_{i+1}$.
If $e$ does not exist in ${\cal DT}_{{\cal R}_i}$, it means
that $e$ is an internal edge of the region retriangulated
when $q$ is inserted in ${\cal DT}_i$. Since
$q$ is a random point in ${\cal R}_i$, the expected number of
such edges is 3 since it equals the average degree of $q$ in ${\cal R}_i$
minus 3.

If $e$ exists  in ${\cal DT}_{{\cal R}_i}$, one end-point $w$ of
$e$ must belong to the disk
of diameter $qv_{i+1}$, denoted $\mbox{disk}[qv_{i+1}]$,
(otherwise any disk through the end-points of $e$ must
contain $q$ or $v_{i+1}$ and $e$ cannot belong to ${\cal DT}_{{\cal R}_i}$).

The expected number of edges of ${\cal DT}_{{\cal R}_i}$ intersecting  $\mbox{disk}[qv_{i+1}]$
is bounded by the sum of the degrees of the vertices in  $\mbox{disk}[qv_{i+1}]$
\begin{eqnarray*}
\lefteqn{
E( \# \{e \in {\cal DT}_{{\cal R}_i} \mbox{ having an end-point}\in {\cal R}_i\cap\mbox{disk }[qv_{i+1}]\})}\\
&=& \frac{1}{|{\cal R}_i|} \sum_{q\in{\cal R}_i} \sum_{w\in {\cal R}_i\cap\mbox{\small disk }[qv_{i+1}]} d^{\circ}_{{\cal DT}_{{\cal R}_i}}(w)\\
&=& \frac{1}{|{\cal R}_i|} \sum_{w\in{\cal R}_i} d^{\circ}_{{\cal DT}_{{\cal R}_i}}(w) \left|\{q\in{\cal R}_i|w\in\mbox{disk }[qv_{i+1}]\}\right|\\
&\leq&  \frac{1}{|{\cal R}_i|} \sum_{w\in{\cal R}_i} d^{\circ}_{{\cal DT}_{{\cal R}_i}}(w) {6}{\alpha}      \hfill\mbox{\footnotesize \hspace*{0.4cm}using Lemma \ref{number_w2}}\\
&\leq& {36}{\alpha}      \hfill\mbox{\footnotesize \hspace*{0.4cm}using the bound of 6 on the average degree of $w$}
\end{eqnarray*}
Notice that Lemma \ref{number_w2} was established for a fixed $w$ and a random $q$ which allows to
use it inside the sum over $w$.
Thus we get a total expected cost for the march bounded by ${36}{\alpha} +3$.
\end{proof}

\begin{lemma} \label{ana_phase3}
The expected number of triangles of ${\cal DT}_i$ visited
during the search for $v_i$ from $t_i$ is $O({\alpha})$.
\end{lemma}
\begin{proof}
All the triangles $t$ examined in phase 3 have a vertex in
the disk of center $q$  passing through $v_{i+1}$.
Thus we can argue similarly as in Lemma \ref{ana_phase2},
denoting $\mbox{disk }|_cqv_{i+1}]$ the disk of center $q$ through
$v_{i+1}$:
\begin{eqnarray*}
\lefteqn{
E( \# \{t \in {\cal DT}_{{\cal R}_i} \mbox{\footnotesize having an end-point}\in \mbox{disk }|_cqv_{i+1}]\})}\\
&\leq& \frac{1}{|{\cal R}_i|} \sum_{q\in{\cal R}_i} \sum_{w\in {\cal R}_i\cap\mbox{\small disk }|_cqv_{i+1}]} d^{\circ}_{{\cal DT}_{{\cal R}_i}}(w)\\
&\leq&  \frac{1}{|{\cal R}_i|} \sum_{w\in{\cal R}_i} d^{\circ}_{{\cal DT}_{{\cal R}_i}}(w) \left|\{q\in{\cal R}_i |w\in{\cal R}_i\cap\mbox{disk }|_cqv_{i+1}]\}\right|\\
&\leq&  \frac{1}{|{\cal R}_i|} \sum_{w\in{\cal R}_i} d^{\circ}_{{\cal DT}_{{\cal R}_i}}(w) {6}{\alpha}      \hfill\mbox{\footnotesize \hspace*{0.4cm}using Lemma \ref{number_w2}}\\
&\leq& {36}{\alpha}      \hfill\mbox{\footnotesize \hspace*{0.4cm}using the bound on the average degree of $w$}
\end{eqnarray*}
\end{proof}

\begin{theorem} \label{insert_time}
The expected cost of inserting $n^{\mbox{th}}$ point in the structure
is $O(\alpha \log_{\alpha} n)$
\end{theorem} 
\begin{proof}
By linearity of expectation, Lemmas \ref{ana_phase1}, \ref{ana_phase2} and \ref{ana_phase3} prove that
the expected cost at one level is  $O({\alpha})$.
Since the expected height of the structure is $\log_{\alpha} n$, we get the claimed result.
(The analysis is similar to the ananlysis for skip lists \cite{mr-ra-95}.)
\end{proof} 

\begin{theorem} \label{construction_time}
The construction of the Delaunay triangulation of a set of $n$ points
is done in expected time $O(\alpha n\log_{\alpha} n)$ and $O(\frac{\alpha}{\alpha-1} n)$ space.
The expectation is on the randomized sampling and the order of insertion,
with no assumption
on the point distribution.
\end{theorem} 
\begin{proof}
Easy corollary of Theorem \ref{insert_time}.
\end{proof}

\section{Tuning parameters\label{tune}}

We have proved that our structure is worst case optimal in
the expected sense for any
set of points.
In this section, we will focus on more practical cases,
and tune the algorithm to be optimal on
random distribution. In that case, many events such as
that a point has high degree and that it is
the nearest neighbor of a random point
can be considered as independent.

\subsection{Phase 1}
\begin{figure} \begin{center} 
\def\IPEfile{Tune1.ipe}\input{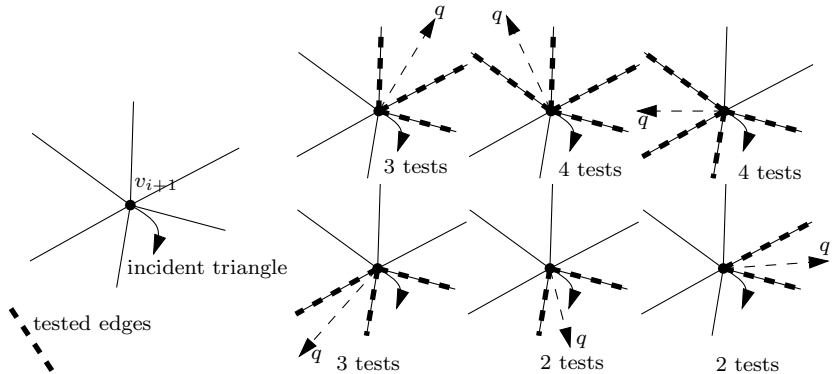}
\caption{\label{Tune1}Number of orientation tests in phase~1}
\end{center} \end{figure}

We can assume that,
$d^{\circ}_{{\cal DT}_{i}}(v_{i+1})=6$
(and not only $\leq 36$ as proved in Lemma \ref{ana_phase1}).
And thus if the turn around $v_{i+1}$ is done in clockwise
or counterclockwise direction depending on the position
of segment $v_{i+1}q$ with respect to the starting triangle,
and assuming that this position is random around $v_{i+1}$
the expected number of orientation tests is 3.
Figure \ref{Tune1} shows the different cases to average,
the edges $v_{i+1}w$ such that an orientation test
$v_{i+1}wq$ is performed are indicated, for a typical degree 6 vertex
in the triangulation.

\subsection{Phase 2}

Bose and Devroye \cite{bd-irgo-95}
proved that the expected number of edges
of a Delaunay triangulation of random points
crossed by a line segment of length $l$ is
$O(l\sqrt{\gamma})$ where $\gamma$ is the point density.
Our experiments shows that the constant is 2.

The expected number of points in the disk of center $q$ passing through
$v_{i+1}$ is $\alpha - 1$. Indeed, if the
points of ${\cal R}_i$ are sorted by increasing distance from $q$,
$v_{i+1}$ is the first point in ${\cal R}_{i+1}$, thus the
number of points in the disk is $k$ with probability
$(1-\frac{1}{\alpha})^k\frac{1}{\alpha}$,
and the expected number is $\frac{1}{\alpha}\sum(1-\frac{1}{\alpha})^k=\alpha-1$.
Thus if $l$ is the length of $qv_{i+1}$
the density of points in ${\cal DT}_i$ is
$\frac{\alpha}{\pi l^2}$.

Thus we conclude that  the expected number of edges
of ${\cal DT}_i$ intersecting segment
$qv_{i+1}$ is $2 l{\sqrt{\frac{\alpha}{\pi l^2}}}=
\frac{2\sqrt{\alpha}}{\sqrt{\pi}}$.

For each edge $ww'$ crossed, two orientation tests are performed:
if $w$ is the newly examined vertex, orientations of
triangles $wqv_{i+1}$ and $qww'$ are computed.

We have to point out, that in the orientation tests of kind
$wqv_{i+1}$, the edge $qv_{i+1}$ remains constant, and
thus some computations do not need to be done for each test.

\subsection{Phase 3\label{tune-3}}

Phase 3 is more difficult to analyze precisely, but a rough bound
is that the number of candidate vertices examined (with shortest distance)
is less than two
and that we examine less than 8 triangles in total.

In fact, we modified phase 3, instead of really searching for
$v_i$, the nearest neighbor of $q$ in ${\cal S}_i$, we just define
$v_i$ as the nearest among the three vertices of $t_i$.
\RRonly{(see Appendix).}
Thus this modified phase 3 reduced to three distance computations
and two comparisons.

\subsection{Tuning $\alpha$}

We will count more precisely the number of operation needed
to evaluate our primitives. More exactly, we count the number
of floating point operations (f.p.o.)
 without making diistinctions between additions,
subtractions or multiplications.

The total evaluation at a given level is
$3+\frac{\sqrt{\alpha}}{\sqrt{\pi}}$ orientation tests involving $qv_{i+1}$,
$\frac{\sqrt{\alpha}}{\sqrt{\pi}}$ other orientation tests and
3 distance computations.

Orientation tests always using points $q$ and $v_{i+1}$
 can be done using 5 f.p.o.\ to initialize plus
4  f.p.o.\ for each test.
Other orientation tests need 7 f.p.o.\ each,
and square distance computations need 5 f.p.o.\ each. 

Thus the total cost in terms of number of f.p.o.\ at level $i$ is

	
\[
5+ 4(3+\frac{\sqrt{\alpha}}{\sqrt{\pi}})+7\frac{\sqrt{\alpha}}{\sqrt{\pi}}+5\cdot 3
= 32 + 6.2 \sqrt{\alpha}.
\]

Since the number of level is $\log_{\alpha}n=\frac{\log_2 n}{\log_2\alpha}$
we get a cost of $c_0(n)=(32 + 6.2\sqrt{\alpha})\left\lceil\frac{\log_2 n}{\log_2\alpha}\right\rceil$
which is close to its minimum ( $\in[13.3\log_2 n, 14\log_2 n]$)
for $\alpha\in[18,90]$, with the minimum occuring for $\alpha\simeq 40$.

\subsection{Comparison with \cite{msz-frplw-96}\label{mix_msz}}

\SoCG{
Similar counting of f.p.o. in M\"ucke et al. algorithm,
using a random sample of $\beta \sqrt[3]{n}$ points,
produces a cost of
\[
c_{MSZ}(n)=5+ 4(3+\frac{ \frac{n}{\beta n^{\frac{1}{3}}} }{\sqrt{\pi}})+7\frac{\frac{n}{\beta n^{\frac{1}{3}}}}{\sqrt{\pi}}+5\beta \sqrt[3]{n}
= 17+ \sqrt[3]{n} \left( \frac{6.2}{\sqrt{\beta}} + 5 \beta \right)
\]
which is close to its minimal value for $0.5<\beta<1$.

\begin{figure} \begin{center} 
\def\IPEfile{Theoretical.ipe}\input{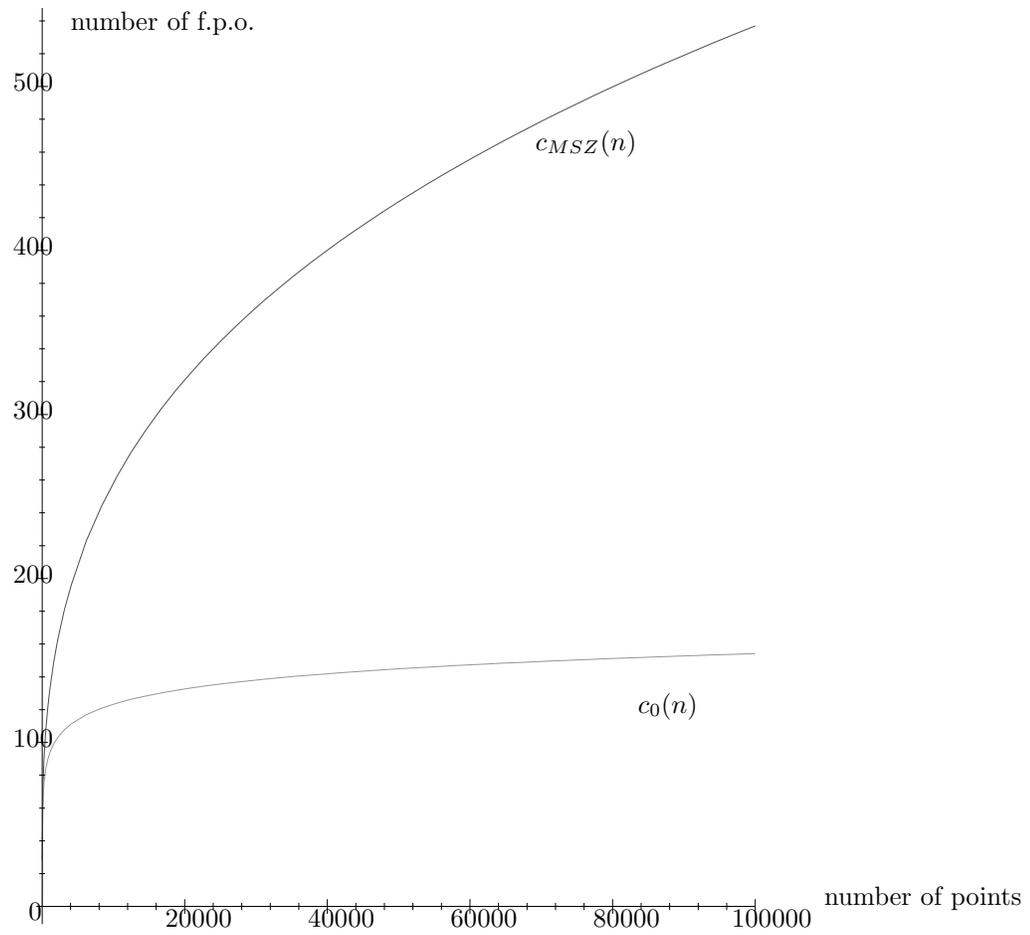}
\caption{\label{Theoretical}Comparison of number of floating point operations
between $c_0(n)$ and $c_{MSZ}(n)$ for $\alpha =40$ and $\beta = 1$.}
\end{center} \end{figure} 

As shown by the comparison of the two curves in Figure \ref{Theoretical},
 our method is potentially
much better than \cite{msz-frplw-96}, even for a small number of points.
However, this method to analyze our approach hides the discontinuity
of the cost, since the effective number of levels is necessarily an integer.
To have a better comprehension of what happens for a small number of points,
we can draw the cost of inserting a point in a structure having a fixed number of
levels.

The classical walk from a random point in the structure costs
\[
c_{walk}(n)=5+ 4(3+\frac{\sqrt{n}}{\sqrt{\pi}})+7\frac{\sqrt{n}}{\sqrt{\pi}}
= 17+6.2 \sqrt{n}
\]
which is also the cost of inserting in our structure up
to the time a second level is created.

When $k$ levels have been created, the cost is
\[
c_k(n)=c_{walk}\left(\frac{n}{\alpha^k}\right)+15 k + k \cdot c_{walk}(\alpha)
\]

We can alternatively mix this multilevel approach with M\"ucke et al's.
sampling at the first level of the structure.
In that case, the cost is
\[
c^{\star}_k(n)=c_{MSZ}\left(\frac{n}{\alpha^k}\right)+15 k + k \cdot c_{walk}(\alpha)
\]
%

\begin{figure} \begin{center} 
\def\IPEfile{Levels.ipe}\input{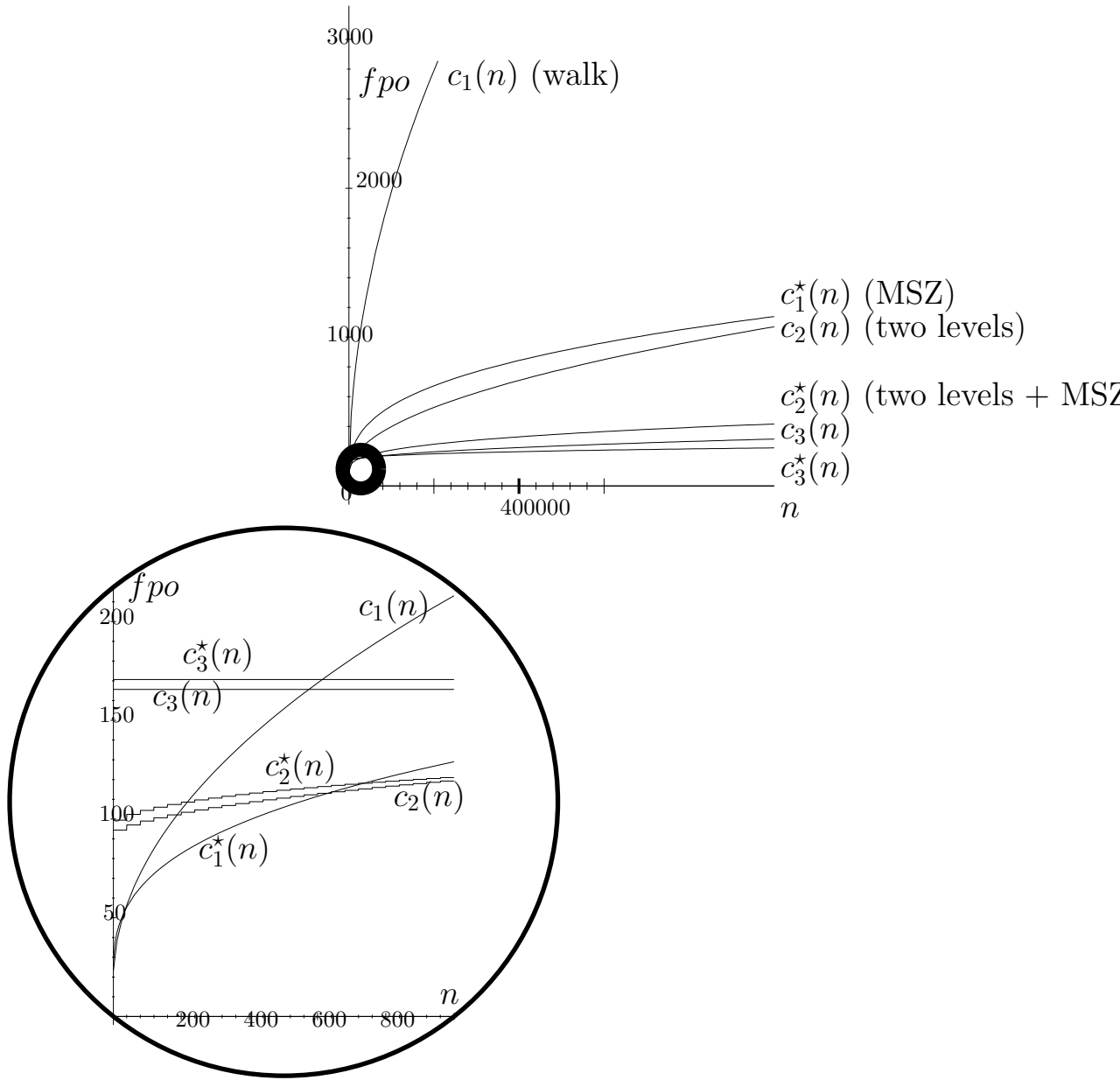}
\caption{\label{Levels}Comparison of number of floating point operations
between $c_k(n)$ and $c^{\star}_k(n)$ for $\alpha =40$.}
\end{center} \end{figure} 

This comparison (see Figure \ref{Levels})
shows that  \cite{msz-frplw-96} ($c^{\star}_1(n)$)
becomes better than the simple march ($c_1(n)$)
for $n>40$.
The two level structure ($c_2(n)$) becomes better than
the single level structure ($c_1(n)$) for $n>180$ and
better than \cite{msz-frplw-96} ($c^{\star}_1(n)$) for $n>600$.
The main information is that the structure presented
in that paper should be significantly better than \cite{msz-frplw-96}
for $10000<n$.
}
\SoCGsubmit{
Details are ommited here (see full version).
A similar analysis of M\"ucke et al. \cite{msz-frplw-96}
allows to conclude that the time complexity should improve
as soon as $n$ is few hundreads, and gain up to a factor of four
for $n=10^6$.
}

\SoCG{
\section{Implementation\label{implement}}

\subsection{Deletion}
The above structure supports insertions and queries as explained
above, but also deletions.
Since there is no complicated data structure to maintain, deletions
can be handled by just deleting the removed point at each level
where it appears.

This can be done in output-sensitive time \cite{c-cdt-87,agss-ltacv-89},
and thus the deletion of a random point
is done in expected constant time since a point
appears at an expected constant number of levels and its
expected degree $k$ is also constant.

From a practical point of view, and to keep the simplicity
of the algorithm, a simpler suboptimal algorithm should be preferred.
It can be done in $O(k^2)$ time, for example by flipping to reduce
the degree of the deleted vertex to 3, and flipping again to restore
the Delaunay property. Another simple algorithm
 consists in finding the Delaunay triangle incident to an edge
of the hole in $O(k)$ time which also yields an $O(k^2)$ time algorithm.
Both algorithms are efficient in practice and needs only few
micro-seconds (about 30 in a random triangulation)
to delete a point once it had been localized.

\subsection{Arithmetic degree}
The algorithm above is designed to make a parsimonious use
of high degree tests \cite{tlp-rpqiv-96}.
More precisely, the location phase uses only 
 orientation tests on three points
in phases 1 and 2, and distance computation and angle
comparisons with $\frac{\pi}{2}$ in phase 3.
All these tests are degree 2 tests.
Clearly, updates need to use in-circle tests which are of degree 4.

An alternative to phase 3 should have to use in-circle tests
to limit the explored triangles in ${\cal DT}_i$ to those
whose circumcircle contains $q$. Such variant may explore
fewer triangles and be easier to analyze, but may use more degree 4 tests.

\subsection{Robustness issues and degeneracies}

Degeneracies are solved by handling special cases:
if two points have the same coordinates, then
the insertion is not done, if four points are cocircular, then
the last point inserted is considered as inside the disk
defined by the others.

We use exact arithmetic for 24 bits integers,
and thus coordinates of our points are integers
in range $[-16777216, 16777216]$ (up to a multiplication
by a power of 2).
Using this restricted kind of data, double precision
computation is exact on degree 2 tests and
 almost never leads to precision problems on degree 4 predicates.
Nevertheless, the exactness of all computations are verified
by an arithmetic filter and exact computation is performed
if needed.

\subsection{Code parameters \label{parameters}}
The following parameters can be specified:
\begin{itemize}
\item maximal number of levels
\item $\alpha$ the ratio between two levels
\item the minimal number of points to use the higher level for point location
\item the minimal number of points to use $MSZ$ sampling at one of the higher levels
\item $\beta$ the constant for the size of $MSZ$ sample.
\end{itemize} 
Our default parameters are 
\begin{itemize}
\item number of levels unlimited
\item $\alpha =30$.
\item  minimal size to use hierarchy is 20.
\item minimal size to use MSZ is 20.
\item $\beta = 1$.
\end{itemize} 
We found that the code is relatively insensitive to
the parameters.
For reasonable changes of these parameters,
(up to a factor 2) the computation time is not greatly
affected.
Using these configuration parameters, our code can be used to run
\begin{itemize}
\item the usual walk algorithm (only one level and minimal size for MSZ=$\infty$),
\item the M\"ucke et al. algorithm \cite{msz-frplw-96} (only one level),
\item the hierarchical algorithm described in this paper (minimal size for MSZ=$\infty$),
\item the mixed method suggested in Section \ref{mix_msz} (default parameters above).
\end{itemize} 

\subsection{Experimental results}

\subsubsection{Data sets}

We claim that our algorithm performs  well on random point sets,
and has acceptable worse case complexity.
To illustrate this fact, we will test it with
the
realistic and degenerate data sets.
For each kind of data, we used sets of size 5,000, 50,000
and 500,000 points.
The coordinates are random on 24 bits and the constraints
such that the points are on a parabola are verified, up to
the rounding arithmetic errors.
\begin{figure} \begin{center}
\def\IPEfile{Examples.ipe}\input{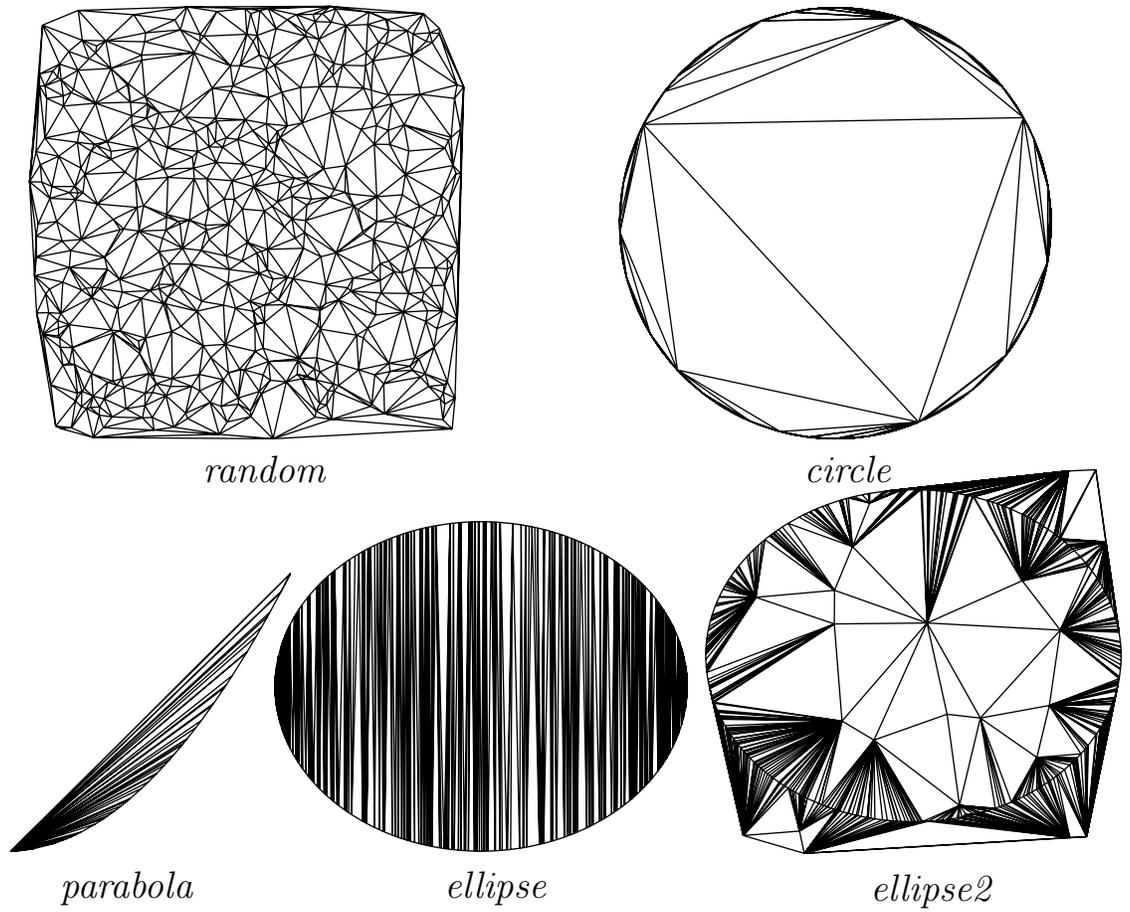} 
\caption{\label{Examples}Data sets.}
\end{center} \end{figure}

\begin{figure*}[t]
\begin{center}
\begin{tabular}{|l|r||r|r|r|r|}
\hline
distribution   	& size	& walk		& \cite{msz-frplw-96}	& hierarchy	& hierarchy + MSZ\\\hline
random          & 5000	& 0.3		& 0.17			& 0.15		& 0.14		\\\hline
random          & 50000	& 12		& 3.8			& 2.7		& 2.3		\\\hline
random          & 500000& 460		& 72			& 36		& 31		\\\hline \hline
ellipse2        & 5000	& 0.53		& 0.34			& 0.21		& 0.20		\\\hline
ellipse2        & 50000	& 49		& 21			& 3.9		& 3.5		\\\hline
ellipse2        & 500000& 930		& 760			& 57		& 49		\\\hline\hline
ellipse         & 5000	& 2.2		& 0.46			& 0.31		& 0.21		\\\hline
ellipse         & 50000	& 187		& 21			& 3.9		& 3.7		\\\hline
ellipse         & 500000& long		& 270			& 54		& 55		\\\hline \hline
parabola        & 5000	& 2.5		& 0.31			& 0.21		& 0.16		\\\hline
parabola        & 50000	& 87		& 5.9			& 3.2		& 3.0		\\\hline
parabola        & 500000& long		& 74			& 69		& 45		\\\hline \hline
circle          & 5000	& 0.15		& 0.13			& 0.13		& 0.14		\\\hline
circle          & 50000	& 2.4		& 2.6			& 2.4		& 2.4		\\\hline
circle          & 500000& 39		& 44			& 36		& 36		\\\hline
\end{tabular}
\end{center} 
\caption{Running times\label{compare MSZ}}
\end{figure*} 

\begin{itemize}
\item {\em random}: points evenly distributed in a square.
\item {\em ellipse}: points evenly distributed on an ellipse.
\item {\em ellipse2}: 95\% points evenly distributed on an ellipse plus
			5\% points evenly distributed in a square.
\item {\em circle}: points evenly distributed on a circle.
\item {\em parabola}: points evenly distributed on a parabola,
\end{itemize}

If the {\em circle} and {\em parabola} examples can be considered as
pathological inputs, the {\em ellipse} and {\em ellipse2} examples are
more realistic, Delaunay triangulation of points
 distributed on a curve occurs in practical applications,
for example in shape reconstruction (see Figure \ref{Examples}).

\subsubsection{Results}

Following results are obtained on  a Sun-Ultra1 200 MHz.
The code is written in C++ and compiled with AT-T compiler with
optimizing options. Time has been obtained with the {\tt clock} command
and is given in seconds. The time which is measured is
just the Delaunay computation; it
 does not take into account the time for input or output.

Figure \ref{compare MSZ} gives the computation times
for execution of the code with the different parameters
described in Section \ref{parameters}.
Since it is the same code, the low level primitives
such as in-circle tests or the walk in the triangulation
are identical and it provides a fair comparison between
the different methods.

The last column is always the fastest method. It is significantly better than MSZ
for very large sets of random points, and the difference is even more
important on data set {\em ellipse2} which is representative of real
applications.

\subsubsection{Comparison with other software}

We have compared with some Delaunay softwares available on the WWW:
\begin{itemize}
\item {\tt qhull} by Bradford Barber and Hannu Huhdanpaa,
  duality with 3D convex hull \cite{bdh-qach-93}
 (available at \\ http://www.geom.umn.edu/locate/qhull).
\item {\tt div-conquer} by Jonathan Shewchuk,
 divide and conquer \cite{s-te2dq-96}
\item {\tt sweep} by Jonathan Shewchuk,
 plane sweep
\item {\tt incremental} by Jonathan Shewchuk,
 incremental with M\"ucke et al. localization.\\
 These three codes supports exact arithmetic on {\tt double}
 (available at\\
 {\small http://www.cs.cmu.edu/$\sim$quake/triangle.research.html}).
\item {\tt Dtree} Delaunay tree structure\cite{bt-rcdt-93} (time includes input)\\
 (available at http://www.inria.fr/prisme/logiciel/del-tree.html).
\item {\tt hierarchy} this paper,  mixed with MSZ.
\end{itemize} 

The execution times in seconds are in Figure \ref{Table other}.
Our method is significantly faster than the other incremental method,
especially in the ellipse cases. Our method is about 50\%
slower than the divide and conquer algorithm.

\begin{figure*}[t]
\begin{center}
\begin{tabular}{|l|r||r|r|r|r|r|r|}
\hline
distribution   	& size	& qhull	& sweep & div-conq	& incr	& Dtree	& hier.	\\\hline\hline
random		& 5000	& 0.65	& 0.21	& 0.11		& 0.29		& 1.4	& 0.14		\\\hline
random		& 50000	& 8.0	& 3.6	& 1.6		& 6.6		& 17	& 2.3		\\\hline
random		& 500000& 101	& 53	& 22		& 150		& swap	& 31		\\\hline\hline
ellipse2	& 5000	& 0.54	& 0.21	& 0.13		& 0.75		& 1.3	& 0.20		\\\hline
ellipse2	& 50000	& 7.8	& 3.2	& 2.16		& 42		& 16	& 3.5		\\\hline
ellipse2	& 500000& 420	& 46	& 29		& 2100		& swap	& 49		\\\hline\hline
ellipse		& 5000	& 0.83	& 0.18	& 0.14		& 2.1		& 1.3	& 0.21		\\\hline
ellipse		& 50000	& 57	& 2.8	& 2.4		& 110		& 14	& 3.7		\\\hline
ellipse		& 500000& swap	& 39	& 33		& 1400		& swap	& 55		\\\hline\hline
parabola	& 5000	& 3.9	& 0.16	& 0.11		& 2.0		& 1.2	& 0.16		\\\hline
parabola	& 50000	& 790	& 2.7	& 2.0		& 110		& 14	& 3.0		\\\hline
parabola	& 500000& swap	& 39	& 28		& 1800		& swap	& 45		\\\hline\hline
circle		& 5000	& 93	& 0.17	& 0.17		& 0.52		& 1.4	& 0.14		\\\hline
circle		& 50000	& 220	& 3.1	& 1.8		& 11		& 15	& 2.4		\\\hline
circle		& 500000& swap	& 22	& 43		& 240		& swap	& 36		\\\hline\hline
\end{tabular} 
\end{center} 
\caption{Comparisons with other softwares\label{Table other}}
\end{figure*} 

}

\section{Conclusion}

We proposed a new hierarchical data structure to compute the Delaunay triangulation
of a set of points in the plane.
It  combines
good worst case randomized complexity,
 fast behavior on real data,
 small memory occupation
and dynamic updates (insertion and deletion of points).

Referring to Su and Drysdale \cite{sd-csdta-97} study of several techniques
and our comparisons with Shewchuk implementation \cite{s-te2dq-96}
of some of these techniques,
we have shown that our implementation is competitive
with other approaches on random data.
Furthermore, we can prove that the performances remains good
on pathological inputs.
Finally, one of the main advantage of this algorithm is
to  allow a dynamic setting.

The main idea of our structure is to perform point location using several levels.
The lowest level just consists of the triangulation, then each
level contains the triangulation of a small sample of the levels below.
Point location is done by marching in a triangulation to determine
the nearest neighbor of the query at that level, then the march restart
from that neighbor at the level below.
Location at highest level is done using \cite{msz-frplw-96} which is efficient
for small set of points.

One characteristics of the structure is that best time performance is obtained
with a ratio of about three per cent between two levels, which yields to
few levels (three or four typically) and a small memory occupation.
The structure is simple and does not need additional features such as buckets.

Such structure can be generalized to other problems.
The two main ingredients of the proofs are bounds on
the maximal degree of the nearest neighbor graph and
the expected degree of a random vertex in the Delaunay triangulation.
The first generalizes well in higher dimension, while the second
becomes an data sensitive parameter (constant for random points,
$n^{\lceil (d-1)/2\rceil}$ in the worst case).
A generalization for computing the trapezoidal map can also be done.

\subsection*{Code}
A demo version compiled for Sun Solaris and SGI
is available at \\
http://www.inria.fr/prisme/logiciels/del-hierarchy/.

\subsection*{Acknowledgement}
The author would like to thank Hervé Brönnimann, Jonathan Shewchuk,
Jack Snoeyink  and Mariette Yvinec
for helpful discussions and careful reading of this paper.

 \small
\bibliography{geom}

\begin{thebibliography}{AGSS89}

\bibitem[AGSS89]{agss-ltacv-89}
A.~Aggarwal, Leonidas~J. Guibas, J.~Saxe, and P.~W. Shor.
\newblock A linear-time algorithm for computing the {Voronoi} diagram of a
  convex polygon.
\newblock {\em Discrete Comput. Geom.}, 4(6):591--604, 1989.

\bibitem[BD95]{bd-irgo-95}
P.~Bose and L.~Devroye.
\newblock Intersections with random geometric objects.
\newblock Technical report, School of Computer Science, McGill University,
  1995.
\newblock Manuscript.

\bibitem[BDH93]{bdh-qach-93}
C.~B. Barber, D.~P. Dobkin, and H.~Huhdanpaa.
\newblock The {Quickhull} algorithm for convex hull.
\newblock Technical Report GCG53, Geometry Center, Univ. of Minnesota, July
  1993.

\bibitem[BT86]{bt-hrodt-86}
Jean-Daniel Boissonnat and Monique Teillaud.
\newblock A hierarchical representation of objects: {The} {Delaunay} tree.
\newblock In {\em Proc. 2nd Annu. ACM Sympos. Comput. Geom.}, pages 260--268,
  1986.

\bibitem[BT93]{bt-rcdt-93}
Jean-Daniel Boissonnat and Monique Teillaud.
\newblock On the randomized construction of the {Delaunay} tree.
\newblock {\em Theoret. Comput. Sci.}, 112:339--354, 1993.

\bibitem[Che86]{c-bvdcp-86}
L.~P. Chew.
\newblock Building {Voronoi} diagrams for convex polygons in linear expected
  time.
\newblock Technical Report PCS-TR90-147, Dept. Math. Comput. Sci., Dartmouth
  College, Hanover, NH, 1986.

\bibitem[Che87]{c-cdt-87}
L.~P. Chew.
\newblock Constrained {Delaunay} triangulations.
\newblock In {\em Proc. 3rd Annu. ACM Sympos. Comput. Geom.}, pages 215--222,
  1987.

\bibitem[DMT92]{dmt-fddtl-92}
Olivier Devillers, Stefan Meiser, and Monique Teillaud.
\newblock Fully dynamic {Delaunay} triangulation in logarithmic expectedtime
  per operation.
\newblock {\em Comput. Geom. Theory Appl.}, 2(2):55--80, 1992.

\bibitem[GKS92]{gks-ricdv-92}
Leonidas~J. Guibas, D.~E. Knuth, and Micha Sharir.
\newblock Randomized incremental construction of {Delaunay} and {Voronoi}
  diagrams.
\newblock {\em Algorithmica}, 7:381--413, 1992.

\bibitem[Law77]{l-scsi-77}
C.~L. Lawson.
\newblock Software for {$C^{1}$} surface interpolation.
\newblock In J.~R. Rice, editor, {\em Math. Software III}, pages 161--194.
  Academic Press, New York, NY, 1977.

\bibitem[MR95]{mr-ra-95}
R.~Motwani and P.~Raghavan.
\newblock {\em Randomized Algorithms}.
\newblock Cambridge University Press, New York, NY, 1995.

\bibitem[MSZ96]{msz-frplw-96}
Ernst~P. M{\"u}cke, Isaac Saias, and Binhai Zhu.
\newblock Fast randomized point location without preprocessing in two- and
  three-dimensional {Delaunay} triangulations.
\newblock In {\em Proc. 12th Annu. ACM Sympos. Comput. Geom.}, pages 274--283,
  1996.

\bibitem[Mul91]{m-rmstd-91}
K.~Mulmuley.
\newblock Randomized multidimensional search trees: {D}ynamic sampling.
\newblock In {\em Proc. 7th Annu. ACM Sympos. Comput. Geom.}, pages 121--131,
  1991.

\bibitem[Mul94]{m-cgitr-93}
K.~Mulmuley.
\newblock {\em Computational Geometry: An Introduction Through Randomized
  Algorithms}.
\newblock Prentice Hall, Englewood Cliffs, NJ, 1994.

\bibitem[PY92]{py-nng-92}
M.~S. Paterson and F.~F. Yao.
\newblock On nearest-neighbor graphs.
\newblock In {\em Proc. 19th Internat. Colloq. Automata Lang. Program.}, volume
  623 of {\em Lecture Notes Comput. Sci.}, pages 416--426. Springer-Verlag,
  1992.

\bibitem[SD97]{sd-csdta-97}
P.~Su and R.~Drysdale.
\newblock A comparison of sequential {Delaunay} triangulation algorithms.
\newblock {\em Comput. Geom. Theory Appl.}, 7:361--386, 1997.

\bibitem[She96]{s-te2dq-96}
J.~R. Shewchuk.
\newblock Triangle: engineering a $2$d quality mesh generator and {Delaunay}
  triangulator.
\newblock In {\em First Workshop on Applied Computational Geometry}.
  Association for Computing Machinery, May 1996.

\bibitem[TLP96]{tlp-rpqiv-96}
R.~Tamassia, G.~Liotta, and F.~P. Preparata.
\newblock Robust proximity queries in implicit {Voronoi} diagrams.
\newblock In {\em Proc. 8th Canad. Conf. Comput. Geom.}, page~1, 1996.

\end{thebibliography}
\bibliographystyle{alpha}

\RRonly{
\appendix
\section{Appendix}
Unfortunately,  Theorem \ref{insert_time} does not hold for the modified
version of $v_i$ suggested at Section \ref{tune-3}.
On Figure \ref{Counterex3}, for all the points marked by a cross, $w_0$
is the nearest neighbor
among the three vertices of the Delaunay triangle containing it,
but  $w_0$ does not have bounded degree.
Thus, with some constant probability $\simeq \frac{1}{\alpha ^3}$
the three vertices of a triangle are in the sample and the point inside is not,
and phase 1 has an non constant cost $\frac{n}{\alpha}$.

We hope that something is still provable!
Anyway, the situations creating problems for the modified
algorithm are fairly pathological.

\begin{figure} \begin{center} \def\IPEfile{Counterex3.ipe}\input{Counterex3.ipe}
\caption{\label{Counterex3}Counter example for optimality of modified phase 3.}
\end{center} \end{figure} 

}

\end{document}